\documentclass[preprint,showpacs,preprintnumbers,amsmath,amssymb]{revtex4}

\usepackage{dcolumn}
\usepackage{bm}
\usepackage{braket}
\usepackage{amsmath}
\usepackage{txfonts}
\usepackage[dvipdfmx]{graphicx}
\usepackage{float}

\begin{document}
\preprint{KUNS2496}
\preprint{KOBE-TH-14-05}

\title{Dynamics of Electroweak Gauge Fields during and after Higgs Inflation}

\author{Ippei Obata$^{1,2}$}
\email{obata@tap.scphys.kyoto-u.ac.jp}
\author{Takashi Miura$^{2}$}
 \email{takashi.miura@people.kobe-u.ac.jp}
\author{Jiro Soda$^{2}$}
 \email{jiro@phys.sci.kobe-u.ac.jp}

\affiliation{
$^{1}$Department of Physics, Kyoto University, Kyoto, 606-8502, Japan\\
$^{2}$Department of Physics, Kobe University, Kobe, 657-8501, Japan
}

\collaboration{CLEO Collaboration}

\date{\today}

\begin{abstract}
We study the dynamics of $SU(2)_L\times U(1)_Y$ electroweak gauge fields during and after Higgs inflation. 
In particular, we investigate configurations of the gauge fields during inflation and find the gauge fields remain topologically non-trivial. 
 We also find that the gauge fields grow due to parametric resonances caused by oscillations of a Higgs field after inflation.
 We show that the Chern-Simons number also grows significantly. 
Interestingly, the parametric amplification gives rise to sizable magnetic fields after the inflation whose final
 amplitudes depend on the anisotropy survived during inflation.
\end{abstract}

\pacs{Valid PACS appear here}
\maketitle

\section{Introduction}

 It is well known that an inflationary scenario has succeeded in solving various cosmological problems of the standard big bang model and predicting scale-invariant fluctuations observed in the large scale structure of the universe and the cosmic microwave background radiation (CMB). In this scenario, a scalar field called  inflaton
is supposed to be a source of quasi-exponential expansion of the universe. 
Thus, it is important to identify the inflaton in a model of particle physics. 
An attractive possibility is that the inflaton is one of scalar fields in a supersymmetric extension of 
the standard model of particle physics. So far, however, there exists no signal of supersymmetry in experiments at the LHC, which has
 been supposed to discover supersymmetric particles. This is one of the reasons people tend to prefer  Higgs inflation 
proposed in \cite{Bezrukov:2007ep}, where the Higgs field plays a role of the inflaton. 
It should be emphasized that inflation can be realized in the electroweak sector of the standard model of particle physics. 
Remarkably, new data released  by Planck strongly support  Higgs inflation~\cite{Ade:2013uln}
although recent BICEP2 results raised a challenge to  Higgs inflation~\cite{Ade:2014xna}.

Needless to say, gauge fields are essential ingredients of electroweak theory.
Thus, it is intriguing to explore the cosmological dynamics of the electroweak gauge fields
in an  inflationary scenario.
Indeed, in view of recent findings concerning roles of gauge fields in inflation ~\cite{Maleknejad:2012fw}, it is worth investigating
a possible role of the electroweak gauge fields in Higgs inflation.   
 It should be noted that the dynamics of electroweak theory coupled with Einstein gravity 
has been already studied in \cite{Emoto:2002fb}. There, it is found that a non-trivial
local minimum appears in the effective potential of the gauge fields.
 In their paper,  however, the presence of a cosmological constant is assumed to realize inflation.
In fact, it is inevitable to consider a concrete inflation model for making predictions
on observables at present. Fortunately, as mentioned above, electroweak theory itself prepares  Higgs inflation. 
Thus, it is worth redoing their analysis in the context of  Higgs inflation.

The other motivation stemmed from primordial magnetic fields~\cite{Grasso:2000wj,Giovannini:2003yn}. 
It is well known that there exist magnetic fields in galaxies. Moreover, there are several observational 
evidences~\cite{Bonafede:2010xg,Tavecchio:2010mk,Neronov:1900zz,Essey:2010nd} for the presence of magnetic fields
 in inter galaxies or inter clusters of galaxies, which seems difficult to explain by astrophysical processes. 
One attractive mechanism for producing magnetic fields with such a long correlation length is
generation of magnetic fields in inflation.
Indeed, the generation of primordial magnetic fields during inflation has been discussed 
in many papers~\cite{Turner:1987bw,Ratra:1991bn,Dolgov:1993vg,
Giovannini:2000dj,Bamba:2003av,Bamba:2006ga,Martin:2007ue}. 
In the course of these studies, strong coupling and backreaction problems are recognized~\cite{Demozzi:2009fu}. 
Although a partial answer to the backreaction problem is given in \cite{Kanno:2009ei} and 
there is a proposal for resolving the strong coupling problem~\cite{Ferreira:2013sqa}, it is fair to say that further works 
including theoretical constraints on the magnitude of magnetic fields from observations~\cite{Shaw:2010ea,Fujita:2014sna}
are necessary. 
At this point, it should be noticed that all of these works have been done in Einstein-Maxwell theory.
However, inflation occurs at the energy scale above the electroweak phase transition.
Hence, it is natural to study generation of primordial magnetic fields in the context of Einstein-electroweak theory.
It is legitimate to expect a simple solution to the above difficulties in this natural setup. 
  
 In this paper, therefore, we focus on the dynamics of the $SU(2)_L\times U(1)_Y$ gauge fields during and
 after Higgs inflation aiming at clarifying the evolution of gauge field configurations and seeking
a novel mechanism for generating primordial magnetic fields. 
 We show that there appear topologically non-trivial configurations of the gauge fields during inflation. 
Moreover, the gauge fields require an anisotropic universe as a framework. Interestingly, it turns out that
the initial anisotropy survives during inflation. Of course, the expansion rate is isotropic in accordance with the cosmic no-hair conjecture. 
We also observe that amplification of the gauge fields after inflation occurs 
due to  parametric resonances induced by oscillations of the Higgs field.
We evaluate  magnetic fields generated by the processes and find that sizable magnetic fields are generated.
Remarkably, the anisotropy survived until the end of inflation affects the final amplitude of magnetic fields. 

 The paper is organized as follows. 
In section 2, we provide the action of Einstein-electroweak theory with a non-minimal coupling 
and derive basic equations for analyzing cosmological dynamics of the gauge fields. 
In section 3, we do the numerical calculation and find parametric resonances. 
We also evaluate the change of the Chern-Simons number due to  parametric resonances. 
In section 4, we show that sizable magnetic fields are generated by  parametric resonances. 
We conclude the paper with summary and outlook in Section 5.

\section{Cosmology in Einstein-Electroweak Theory}

In this section, we present basic formulas for studying cosmology in non-minimally coupled Einstein-electroweak theory where
a Higgs field $\Phi$ plays a role of an inflaton.
The action reads 
\begin{gather}
S=\int d^4x\sqrt{-g}\left[\dfrac{M_{pl}^2}{2}R+\xi\Phi^{\dagger}\Phi R-\dfrac{1}{4}F^a_{\mu\nu}F^{a\mu\nu}-\dfrac{1}{4}G^a_{\mu\nu}G^{a\mu\nu}-(D_\mu\Phi)^\dagger(D^\mu\Phi) -\lambda\left(\Phi^\dagger\Phi-\dfrac{v_0^2}{2}\right)^2\right] \ ,
\end{gather}
where $M_{pl}$ is the reduced Planck mass,
$g$ is a determinant of a metric $g_{\mu\nu}$, and $\xi$, $\lambda$ and $v_0$ are parameters. Here, a gauge covariant derivative is defined as
\begin{eqnarray}
D_\mu \Phi = \left(\partial_\mu - i \dfrac{g}{2}\tau^aA^a_\mu-i\dfrac{g'}{2}B_\mu\right)\Phi \ ,
\end{eqnarray}
where $g$ and $g'$ are gauge coupling constants, $A^a_\mu$ and $B_\mu$ are $SU(2)_L\times U(1)_Y$ gauge fields, and $\tau^a$ are Pauli matrices.
Note that the coupling constant $g$ is not related with the metric.
We also defined the $SU(2)_L$ field strength 
 $F^a_{\mu\nu}= \partial_\mu A_\nu^a - \partial_\nu A_\mu^a +g f^{abc} A_\mu^b A_\nu^c $ with  structure constants $f^{abc}$
and the $U(1)_Y$ field strength $G_{\mu\nu} = \partial_\mu B_\nu -\partial_\nu B_\mu$.
Note that, for the parameter region $1 \ll \sqrt{\xi} \lll 10^{17}$, inflation can be realized
in the context of electroweak theory~\cite{Bezrukov:2007ep}.

 Hereafter, we assume a homogeneous universe with a spatial topology of $S^3$. In this case, we can use the Maurer-Cartan invariant one-form basis
 $\sigma^j \ (j=1,2,3)$ which satisfy
\begin{equation}
d\sigma^i=\epsilon^{ijk}\sigma^j\wedge\sigma^k \ , 
\end{equation}
where $\epsilon^{ijk}$ is the Levi-Civita symbol. 
Using these one-forms, the metric can be written as
\begin{align}
ds^2 &= -N^2dt^2+a_1^2((\sigma^1)^2+(\sigma^2)^2)+a_3^2(\sigma^3)^2 \\
       &\equiv -(e^0)^2+(e^1)^2+(e^2)^2+(e^3)^2,
\end{align}
where $N$ is the lapse function and {$e^\mu$  $(\mu=0,1,2,3)$} is the orthonormal basis. 
Here, scale factors $a_1(t)$ and $a_3(t)$ are different in general because of the presence of the $U(1)_Y$ gauge field which is supposed to
have a specific direction
\begin{eqnarray}
{\bf B} =h(t) \sigma^3 \ .
\end{eqnarray}
This anisotropic ansatz is inevitable when we consider cosmology in electroweak theory.
 It is also necessary when a non-trivial coupling between an inflaton and gauge fields exists as in electroweak theory, 
because there is a possibility to have  anisotropic inflation~\cite{Watanabe:2009ct}.
Indeed, we will see Higgs inflation can be regarded as a kind of anisotropic inflation in the sense that
the survived anisotropy is relevant to the gauge field dynamics.
Next, using the Gauge degrees of freedom, we transform the Higgs field
\begin{equation}
\Phi=\dfrac{v}{\sqrt{2}}\begin{pmatrix}
                                           x^2+ix^1 \\
                                           x^4-ix^3 \\
                               \end{pmatrix} \\
\end{equation} 
to
\begin{equation}
\Phi\rightarrow H^{-1}\Phi=\dfrac{1}{\sqrt{2}}\begin{pmatrix}
                                           0 \\
                                           v \\
                               \end{pmatrix} \ , \\
\end{equation}
where $H=x^4+ix^i\tau^i  \ (i=1,2,3)$, with a relation $(x^i)^2+(x^4)^2=1$, is an element of $SU(2)_L$ group. 
Here, $\tau^i$ are Pauli matrices. 
Under this transformation, the trivial gauge field becomes
\begin{equation}
-iH^{-1}dH = \sigma^j\tau^j \ . \label{eq:maurer}
\end{equation}
 Thus, we use the following ansatz of the Higgs fields
\begin{align}
\Phi =\dfrac{1}{\sqrt{2}}\begin{pmatrix}
                                           0 \\
                                           v(t) \\
                               \end{pmatrix} \ . \label{ansatz-higgs}
\end{align}
Finally, we can assume the gauge fields as follows
\begin{eqnarray}
{\mathbf A} =\dfrac{1}{2g}\left[ f_1(t)( \sigma^1\tau^1+\sigma^2\tau^2 )+f_3(t) \sigma^3\tau^3  \right] \ , 
\label{ansatz-gauge}
\end{eqnarray}
where we have taken into account anisotropy due to the $U(1)$ gauge field $B_\mu$.
Substituting the ansatz (\ref{ansatz-gauge}) into the field strength of the $SU(2)_L$ gauge field, ${\bf F} = d {\bf A}-ig {\bf A} \wedge {\bf A} $,
we obtain
\begin{eqnarray}
{\bf F}
&=&\dfrac{\dot{f_1}}{ga_1N}e^0\wedge\left(e^1\dfrac{\tau^1}{2}
+e^2\dfrac{\tau^2}{2}\right)+\dfrac{\dot{f_3}}{ga_3N}e^0\wedge e^3\dfrac{\tau^3}{2} \notag \\
&& \hspace{2cm} + \dfrac{f_1(f_3+2)}{ga_1a_3}\left(e^2\wedge e^3\dfrac{\tau^1}{2}+e^3\wedge e^1\dfrac{\tau^2}{2}\right)+\dfrac{f_1^2+2f_3}{ga_1^2}e^1\wedge e^2\dfrac{\tau^3}{2}  \ .
\end{eqnarray}
It is easy to see a trivial configuration $f_1 = f_3 =0$ has the
field strength ${\bf F}=0$. Moreover,
the configuration with $f_1=f_3=-2$ also gives rise to ${\bf F}=0$.
Indeed, this is a case of pure gauge
\begin{eqnarray}
{\bf A}=-\frac{1}{g}\sigma^j\tau^j=\frac{i}{g}H^{-1}d\mathcal{H}  \ .
\end{eqnarray}
The field strength of the $U(1)_Y$ gauge ${\bf G} =d{\bf B}$ is given by
\begin{equation}
{\bf G} 
=\dfrac{\dot{h}}{Na_3}e^0\wedge e^3+\dfrac{2h}{a_1^2}e^1\wedge e^2.
\end{equation}
Note that $h =  0$ yields ${\bf G}=0$.

 Using the ansatz (\ref{ansatz-higgs}), the term proportional to the curvature  becomes
\begin{eqnarray}
\frac{M_{pl}^2}{2}\left(1+\frac{\xi v^2}{M_{pl}^2}\right)R \ .
\end{eqnarray}
In order to transform the metric into the Einstein frame, we need the conformal transformation
\begin{equation}
\hat{g}_{\mu\nu}=\Omega^2g_{\mu\nu}, \ \Omega\equiv\sqrt{1+\dfrac{\xi v^2}{M_{pl}^2}} \ .
\end{equation}
 Substituting these ansatzes into the action after the conformal transformation, we get the following action in the Einstein frame
\begin{gather}
S=2\pi^2\int dt a_1^2a_3\dfrac{M_{pl}^2}{2}\left[\dfrac{4}{a_1}\dfrac{d}{dt}\left(\dfrac{\dot{a_1}}{N}\right)+\dfrac{2}{a_3}\dfrac{d}{dt}\left(\dfrac{\dot{a_3}}{N}\right)+\dfrac{2}{N}\left(\dfrac{\dot{a_1}}{a_1}\right)^2+\dfrac{4}{N}\left(\dfrac{\dot{a_1}}{a_1}\right)\left(\dfrac{\dot{a_3}}{a_3}\right)+N\left(\dfrac{8}{a_1^2}-\dfrac{2a_3^2}{a_1^4}\right)\right] \notag \\
+2\pi^2\int dt N a_1^2a_3\left[\dfrac{1}{2g^2N^2}\left(\dfrac{2\dot{f_1}^2}{a_1^2}+\dfrac{\dot{f_3}^2}{a_3^2}\right)+\dfrac{\dot{h}^2}{2N^2a_3^2}+\dfrac{1}{2N^2}\left(1+\dfrac{6\xi^2v^2}{M_{pl}^2\Omega^2}\right)\dfrac{\dot{v}^2}{\Omega^2}\right] \notag \\
-2\pi^2\int dt N a_1^2a_3V(f_1,f_3,h,v,a_1,a_3)   \ , 
\end{gather}
where the potential function reads
\begin{eqnarray}
V(f_1,f_3,h,v,a_1,a_3)=\dfrac{\lambda}{4\Omega^4}(v^2-v_0^2)^2+\dfrac{v^2}{8\Omega^2}\left(\dfrac{2f_1^2}{a_1^2}+\dfrac{(f_3-g'h)^2}{a_3^2}\right) \notag 
\\
+\dfrac{1}{2g^2}\left(\dfrac{2f_1^2(f_3+2)^2}{a_1^2a_3^2}+\dfrac{(f_1^2+2f_3)^2}{a_1^4}\right)+\dfrac{2h^2}{a_1^4} \ . \label{eq:pot}
\end{eqnarray}
We can envisage the dynamics of gauge fields based on the potential \eqref{eq:pot}. It consists of three parts: the first term coming from the Higgs potential is independent of the scale factors, the second term is proportional to the inverse of quadratic of scale factors, and the third and fourth term are proportional to the inverse of quartic of the scale factors. In the course of expansion of the universe, 
the terms depending on $a_1, a_3$ become negligible compared to the first term. However, in the early stage of the universe, 
since the second term contains the factor $(f_3-g'h)$, there exists a flat direction of the potential. 
So, the gauge fields first roll down to the bottom of the flat direction where the topology of the gauge fields is non-trivial.
This makes the difference as we will see in the following sections.  

Now, we can derive basic equations.
It is convenient to define new variables as
\begin{equation}
a_1\equiv e^{\alpha+\beta}, \qquad a_3\equiv e^{\alpha-2\beta} 
\end{equation}
where $\alpha$ and $\beta$ describe the average and the anisotropic expansion of the universe, respectively.
In this paper, we use a length unit normalized by GeV$^{-1}$.
Taking the variation with respect to the lapse function $N$, we obtain the Hamiltonian constraint
\begin{gather}
3(\dot{\alpha}^2-\dot{\beta}^2)+e^{-2(\alpha+\beta)}(4-e^{-6\beta})=\dfrac{1}{M_{pl}^2}\left[\dfrac{1}{2g^2}\left(2\dot{f_1}^2e^{-2(\alpha+\beta)}+\dot{f_3}^2e^{-2(\alpha-2\beta)}\right)+\dfrac{\dot{h}^2}{2}e^{-2(\alpha-2\beta)} \right. \notag \\
+\dfrac{1}{2}\left(1+\dfrac{6\xi^2v^2}{M_{pl}^2\Omega^2}\right)\dfrac{\dot{v}^2}{\Omega^2}+\dfrac{\lambda}{4\Omega^4}(v^2-v_0^2)^2+\dfrac{v^2}{8\Omega^2}\left(2f_1^2e^{-2(\alpha+\beta)}+(f_3-g'h)^2e^{-2(\alpha-2\beta)}\right) \notag \\
\left. +\dfrac{1}{2g^2}\left(2f_1^2(f_3+2)^2e^{-2(2\alpha-\beta)}+(f_1^2+2f_3)^2e^{-4(\alpha+\beta)}\right)+2h^2e^{-4(\alpha+\beta)}\right] \ . \label{eq:constraint}
\end{gather}
Hereafter, we can set $N=1$. Other components of Einstein equations can be obtained by taking the variations of the action with respect to $\alpha$ and $\beta$.
The equation for $\alpha$ reads
\begin{align}
&6\ddot{\alpha}+9(\dot{\alpha}^2+\dot{\beta}^2) +(4-e^{-6\beta})e^{-2(\alpha +\beta)} \notag \\
& \qquad =-\dfrac{1}{M_{pl}^2}\Biggl[\dfrac{e^{-2\alpha}}{2g^2}(2\dot{f_1}^2e^{-2\beta}+\dot{f_3}^2e^{4\beta})+\dfrac{\dot{h}^2e^{-2(\alpha -2\beta)}}{2}+\dfrac{3}{2}\left(1+\dfrac{6\xi^2v^2}{M_{pl}^2\Omega^2}\right)\dfrac{\dot{v}^2}{\Omega^2} \notag \\
&\hspace{35mm}-\dfrac{3\lambda}{4\Omega^4}(v^2-v_0^2)^2-\dfrac{v^2e^{-2\alpha}}{8\Omega^2}(2f_1^2e^{-2\beta}+(f_3-g'h)^2e^{4\beta}) \notag \\
&\hspace{35mm}+\dfrac{e^{-4\alpha}}{2g^2}(2f_1^2(f_3+2)^2e^{2\beta}+(f_1^2+2f_3)^2e^{-4\beta})+2h^2e^{-4(\alpha +\beta)}\Biggr] \ .
\end{align}
Similarly, the equation for $\beta$ is given by
\begin{align}
&3\ddot{\beta}+9\dot{\alpha}\dot{\beta}+4(1-e^{-6\beta})e^{-2(\alpha +\beta)} \notag \\
&\qquad =-\dfrac{1}{M_{pl}^2}\Biggl[\dfrac{e^{-2\alpha}}{g^2}(\dot{f_1}^2e^{-2\beta}-\dot{f_3}^2e^{4\beta})-\dot{h}^2e^{-2(\alpha -2\beta)}-\dfrac{v^2e^{-2\alpha}}{4\Omega^2}(f_1^2e^{-2\beta}-(f_3-g'h)^2e^{4\beta}) \notag \\
&\hspace{50mm}+\dfrac{e^{-4\alpha}}{g^2}(f_1^2(f_3+2)^2e^{2\beta}-(f_1^2+2f_3)^2e^{-4\beta})-4h^2e^{-4(\alpha +\beta)}\Biggr] \ .
\end{align}
The equation for the Higgs field can be deduced as
\begin{eqnarray}
&&\ddot{v}+3\dot{\alpha}\dot{v}-\dfrac{\xi v\dot{v}^2}{M_{pl}^2\Omega^2} \nonumber\\
&& \qquad = -\dfrac{v}{1+\dfrac{6\xi^2v^2}{M_{pl}^2\Omega^2}}\left[\left(\dfrac{1}{4}\left(2f_1^2e^{-2(\alpha+\beta)}+(f_3-g'h)^2e^{-2(\alpha-2\beta)}\right)    +\dfrac{6\xi^2\dot{v}^2}{M_{pl}^2\Omega^2}\right)\left(1-\dfrac{\xi v^2}{M_{pl}^2\Omega^2}\right) \right. \nonumber\\
&&  \left. \hspace{45mm}+\dfrac{\lambda(v^2-v_0^2)}{\Omega^2}\left(1-\dfrac{\xi}{M_{pl}^2\Omega^2}(v^2-v_0^2)\right)\right]  \ .
\end{eqnarray}
Moreover, we need the equations for $SU (2)$ gauge fields 
\begin{equation}
\ddot{f_1}+(\dot{\alpha}-2\dot{\beta})\dot{f_1}=-f_1\left(e^{-2(\alpha-2\beta)}(2+f_3)^2+e^{-2(\alpha+\beta)}(f_1^2+2f_3)+\dfrac{g^2}{4}\dfrac{v^2}{\Omega^2}\right) \label{eq:f_1}
\end{equation}
and
\begin{equation}
\ddot{f_3}+(\dot{\alpha}+4\dot{\beta})\dot{f_3}=-2e^{-2(\alpha+4\beta)}(f_1^2+2f_3)-2e^{-2(\alpha+\beta)}f_1^2(f_3+2)-\dfrac{g^2}{4}\dfrac{v^2}{\Omega^2}(f_3-g'h) \label{eq:f_3} \ .
\end{equation}
Finally, the equation for the $U(1)_Y$ gauge field reads
\begin{equation}
\ddot{h}+(\dot{\alpha}+4\dot{\beta})\dot{h}=-4e^{-2(\alpha+4\beta)}h+\dfrac{g'}{4}\dfrac{v^2}{\Omega^2}(f_3-g'h) \label{eq:h},
\end{equation}

Now that we have obtained basic equations, we can move on to the analysis of cosmological dynamics of gauge fields.

\section{Cosmological Dynamics of Gauge Fields}

In this section, we focus on the dynamics of the gauge fields. We will see gauge fields remain non-trivial during inflation
although the energy density of them rapidly decays during inflation in agreement with the cosmic 
no-hair conjecture. However, after inflation, the gauge fields show the parametric resonances due to oscillations of the Higgs field.
Furthermore, we explicitly evaluate the Chern-Simon number to characterize the topology of the gauge fields.

 In the subsequent calculations, we will use the following numerical values adopted from experiments
\begin{equation}
g=0.653, \ g'=0.358, \ v_0=246[\text{GeV}], \ m_H=126[\text{GeV}], 
\ M_{pl}= 2\times10^{18} [\text{GeV}] \ .
\end{equation}
We also set initial values of time derivative of the following variables to be zero for convenience:
\begin{equation}
\dot{f_{1i}}=\dot{f_{3i}}=\dot{h_i}=\dot{v_i}=\dot{\beta}_i=0 \ ,
\end{equation}
where the index $i$ represents the value at the initial time $t=0$. 
According to \cite{Bezrukov:2007ep}, we need to take $\xi\sim\sqrt{\lambda/3}N_e /(0.027)^2$
for inflation to be realized, where $N_e$ is the total $e$-folding number.  Taking into account the relation $\lambda=m_H^2/2v_0^2$, 
 we have chosen $\xi=2\times10^4$.  We also fixed the initial value of the Higgs field as $v_i=10^{17}[\text{GeV}]$. 
As to the metric variables $\alpha$ and $\beta$, we left them arbitrary parameters as long as the positivity of the Hubble squared is guaranteed in Eq. \eqref{eq:constraint}.

\begin{figure}[here]
\centering
\includegraphics[width=10cm,keepaspectratio]{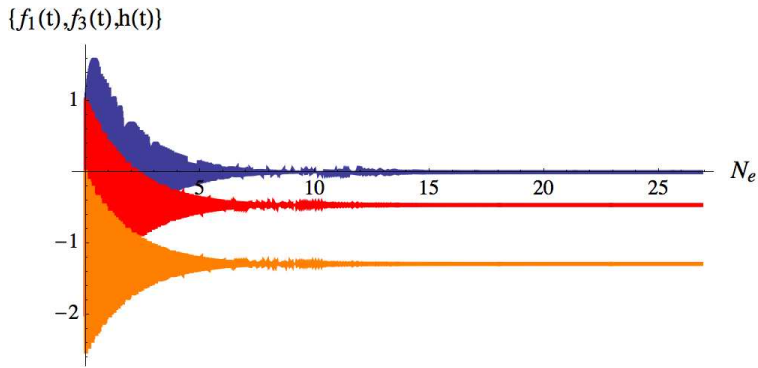}
\caption{The time evolution of $f_1(t)$ (blue), $f_3(t)$ (red), and $h(t)$ (orange) during inflation with the initial conditions
$f_{1i}=-2 \ , f_{3i}=-2 \ , h_{i}=0 \ , \alpha_i=-20 \ , \beta_i=0$. They oscillate near the bottom of the effective potential with damping.}
\label{fig: $f_1f_3h$}
\end{figure}%

 In Fig.\ref{fig: $f_1f_3h$},  we plotted the time evolution of the gauge fields during inflation.
For this calculation, we started from the pure gauge configuration and did not consider the initial anisotropy: 
\begin{equation}
f_{1i}=-2 \ ,\quad f_{3i}=-2 \ ,\quad h_{i}=0 \ ,\quad \beta_{i}=0 \ .
\end{equation}
We chose the initial value $\alpha_i=-20$ and determined $\dot{\alpha}_i$ from the constraint  \eqref{eq:constraint}.  We can see that the gauge fields
 oscillate initially and converge into some values which do not correspond to the pure gauge ones. The similar behavior is also observed in Ref.\cite{Emoto:2002fb}.  

\begin{figure}[here]
\centering
\includegraphics[width=10cm,clip]{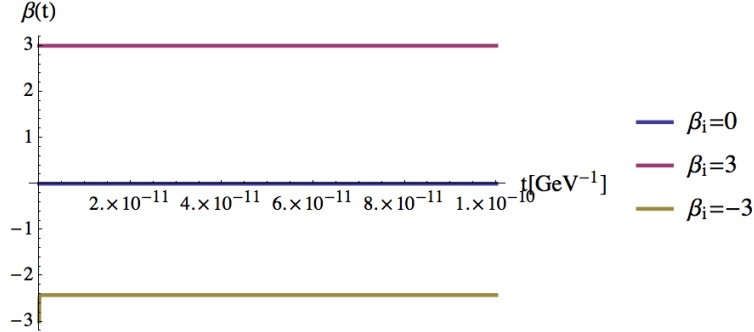}
\caption{The time evolution of $\beta (t)$ for several $\beta_i$ values with the initial conditions $f_{1i}=-2, f_{3i}=-2, h_{i}=0, \alpha_i=-20$.}
\label{fig: $beta$}
\end{figure}%

In Fig.\ref{fig: $beta$}, we plotted $\beta$ for several initial values $\beta_i$.
It turns out the anisotropy remains during inflation although the expansion rate is isotropic. This anisotropy has impact on the evolution of gauge field configurations after inflation because Eqs. (\ref{eq:f_1}), (\ref{eq:f_3}), and  (\ref{eq:h}) explicitly depends on  the anisotropy $\beta$.
Because of this anisotropy as well as the non-trivial gauge fields, Higgs inflation is anisotropic. 

\begin{figure}[here]
\centering
\includegraphics[width=10cm,clip]{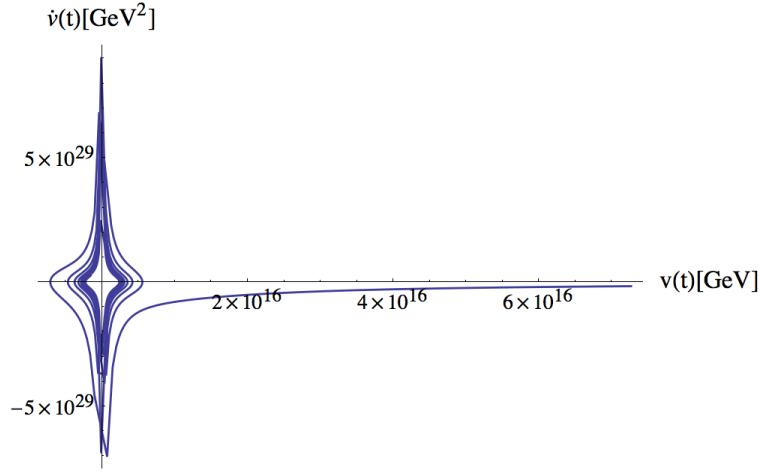}
\caption{The time evolution of the Higgs field in phase space with the initial conditions
 $f_{1i}=-2 \ , f_{3i}=-2 \ , h_{i}=0 \ , \alpha_i=-20 \ , \beta_i=0 $.}
\label{fig: $v$}
\end{figure}

We plotted the time evolutions of the Higgs field in phase space in Fig.\ref{fig: $v$}. In this case, we have about 40 $e$-folding number.
We can see that the Higgs field shows damping oscillation after inflation. During this stage, the gauge field also shows chaotic oscillation.
This kind of chaotic behavior is also found in other models~\cite{Murata:2011wv}
where a gauge kinetic function keeps gauge fields non-zero. While, in this case, the
gauge fields remain non-zero due to the flat direction.

\begin{figure}[here]
\centering
\includegraphics[width=12cm,height=10cm,keepaspectratio]{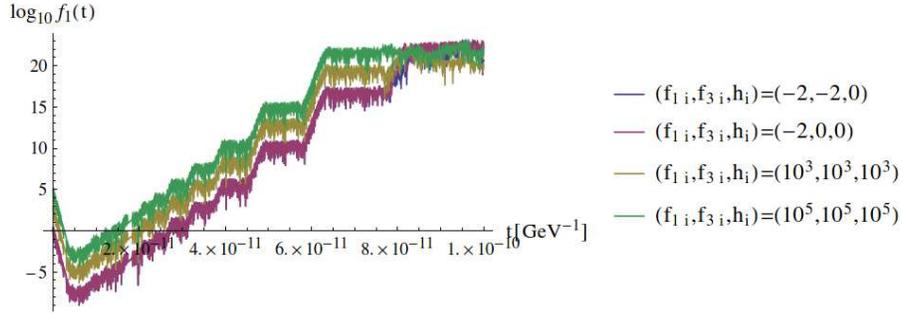}
\caption{The time evolution of $f_1(t)$ after inflation for several initial sets of gauge values with the initial conditions
 $\alpha_i=-20 \ , \beta_i=0$. 
 The final amplitude seems to converge to a value.}
\label{fig: $logf1-2$}
\end{figure}

\begin{figure}[here]
\centering
\includegraphics[width=12cm,height=10cm,keepaspectratio]{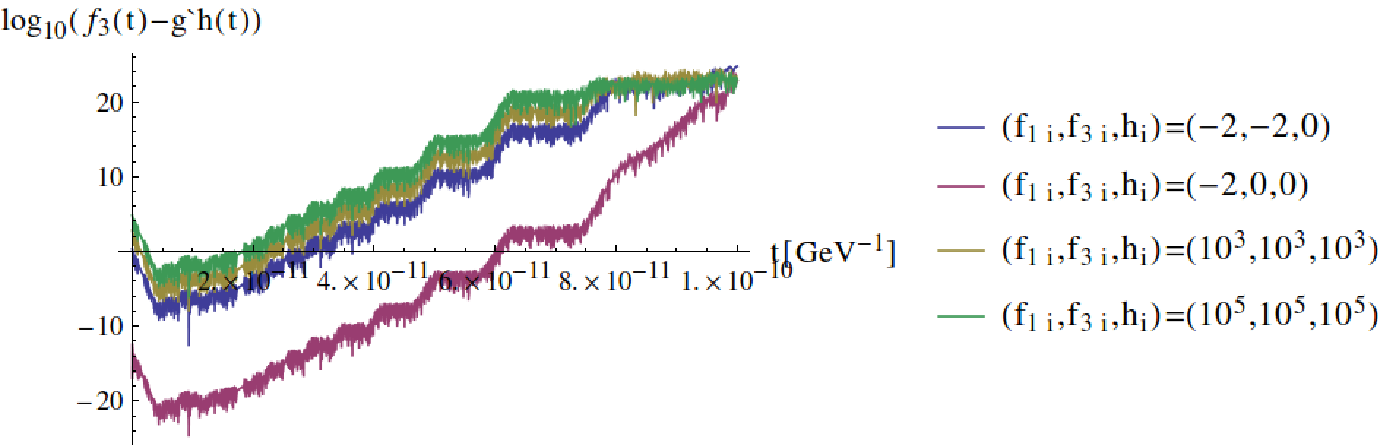}
\caption{The time evolution of $f_3(t)-g'h(t)$ after inflation for several initial sets of gauge values with
 the initial conditions $\alpha_i=-20 \ , \beta_i=0$. }
\label{fig: $logf3-2$}
\end{figure}

In Fig.\ref{fig: $logf1-2$} and Fig.\ref{fig: $logf3-2$}, we plotted the absolute value of gauge fields with different initial conditions. 
There, we can see rapid growth of the amplitude of the gauge fields
for both $f_1$ and $f_3 -g' h$. It is easy to check the energy density of gauge fields also rapidly grows.
Remarkably, the gauge fields converge to a value due to the non-linear effect.
From Fig.\ref{fig: $logf3-2$}, we see even if we started with $f_3 =h=0$, the final value is similar to other cases.

We can understand the behavior in Fig.\ref{fig: $logf1-2$} and Fig.\ref{fig: $logf3-2$} as follows. 
In fact, after inflation, we can approximate the equations for the gauge fields \eqref{eq:f_1}$-$\eqref{eq:h} as follows
\begin{equation}
\ddot{f_1}+\dfrac{g^2}{4}v^2f_1\backsimeq 0    \label{eq:f_1para}
\end{equation}
and
\begin{equation}
(\ddot{f_3}-g'\ddot{h})+\dfrac{g^2+g'^2}{4}v^2(f_3-g'h)\backsimeq 0 \ . \label{eq:f_3para}
\end{equation}
In this era, the Higgs field can be also approximated as
\begin{equation}
v^2\equiv V(t)^2\cos{\omega t}^2=\dfrac{V(t)^2}{2}(1+\cos{2\omega t}) \ , \ V(t)\equiv V_{0}e^{-At} \ ,
\end{equation}
where the parameters $\omega, V_{0}, A > 0$ are determined from numerical calculations.
 We have numerically  verified the condition $A\delta t \ll 1$ is satisfied.
From Eqs. \eqref{eq:f_1para} and \eqref{eq:f_3para}, 
we see a parametric resonance appears in a certain band around the center value 
  where the following relationship between the frequency and amplitude holds; 
\begin{equation}
2\omega=2\dfrac{V(t)}{\sqrt{2}n}  
\ . \label{eq:n0}
\end{equation}
Here, $n$ is an integer and hence there are many resonance bands.
 It is known that, for smaller $n$, the resonance band becomes broad and the growth rate becomes large. 
It is useful to rewrite the above relation as
\begin{equation}
  n=\dfrac{V_{0}}{\sqrt{2}\omega}e^{-At} 
\ . \label{eq:n}
\end{equation}
Since the $V(t)$ is decaying, this condition will be satisfied at a moment which is a center of a band  and a resonance will occur
 for some period around the center of the resonance. 
 As time goes by, the resonance becomes rare 
but the growth rate of amplitude comes to increase when a resonance happens. 
This is because the factor $e^{-At}$ of $V(t)$ 
in the condition \eqref{eq:n} changes slowly  and $n$ gets small as $t$ increases. 
You can see these features in Fig.\ref{fig: $logf1-2$} and Fig.\ref{fig: $logf3-2$}.

\begin{figure}[here]
\centering
\includegraphics[width=12cm,height=10cm,keepaspectratio]{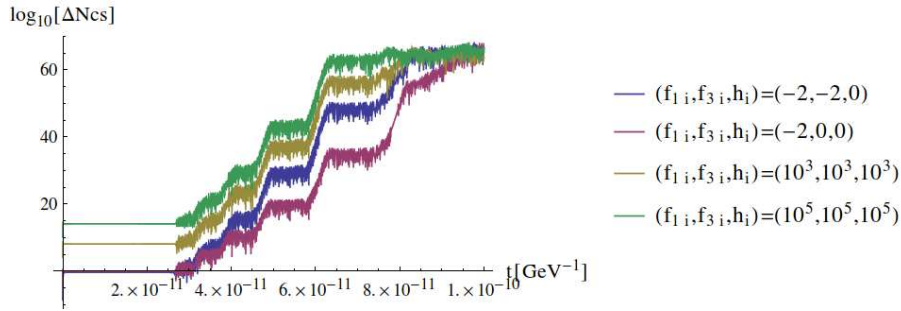}
\caption{The time evolution of $\Delta N_{CS}$ after inflation for several initial sets of gauge values with the initial conditions
 $\alpha_i=-20, \beta_i=0$.}
\label{fig: $logNcs-2$}
\end{figure}%

 Now, we consider the topological configuration of the gauge fields. As you can see from their dynamics, the gauge field configurations are
 quite complicated due to their interaction with the Higgs field.  
Here, let us characterize the gauge filed configurations by the change of Chern-Simons number: 
\begin{eqnarray}
\Delta N_{CS}&\equiv& \dfrac{g^2}{16\pi^2}\int^{t}_{0}dt\sqrt{-g}\text{Tr}(F_{\mu\nu}\tilde{F}^{\mu\nu}) \nonumber\\
&=&-\dfrac{1}{8}\int^{t}_{0} \frac{d}{dt}\left[ f_1^2 f_3 +2f_1^2 + f_3^2 \right]dt   \ .
\end{eqnarray}
We plotted its value as a function of time in Fig.\ref{fig: $logNcs-2$}.
Apparently, the Chern-Simons number is increasing in time. Interestingly, there is a plateau for some period.
Since the presence of the Chern-Simons number indicates CP violation, this result may be relevant to the origin of the baryon asymmetry.

The most important implication of our findings is that the gauge fields are non-trivial 
even at the background level. This would have impact on simulations of preheating~\cite{GarciaBellido:2003wd,DiazGil:2005qp,GarciaBellido:2008ab,Deskins:2013dwa}
where trivial gauge fields are assumed as initial conditions. 

\section{Generation of Magnetic Fields}

In the previous section, we find the growth of gauge fields due to parametric resonances.
In this section, we evaluate the magnitude of primordial magnetic fields produced in this way.
Since the magnetic field is rapidly oscillating, we do not say this is directly transfered to observed magnetic fields.
However, we would like to point out that it is possible to generate sizable magnetic fields with helical structure 
in the presence of this background because of the non-trivial Chern-Simons number. 

 Because of the non-trivial expectation value of the Higgs field, the $SU(2)_L\times U(1)_Y$ gauge symmetry breaks
 down to the $U(1)_{em}$ gauge symmetry. Its gauge potential $A_{em}$ is given by 
the linear combination of $A_\mu^3$ and $B_\mu$  with the Weinberg-angle $\theta_W=\tan^{-1}(g'/g)$, 
\begin{align}
A_{em}&=A^3\sin{\theta_W}+B\cos{\theta_W} \\
  &=\dfrac{1}{\sqrt{g^2+g'^2}}\left( g'\dfrac{f_3}{g}+gh \right) \sigma^3\equiv h_{em}\sigma^3.
\end{align}
Then, its field strength can be calculated as
\begin{equation}
F_{em}=dA_{em}=\dfrac{\dot{h}_{em}}{a_3N}e^0\wedge e^3+\dfrac{2h_{em}}{a_1^2}e^1\wedge e^2 \ .
\end{equation}
So, the magnetic field points in the $e^3$ direction: 
\begin{equation}
B_3=\dfrac{2h_{em}}{a_1^2} \label{eq:B3}.
\end{equation}

\begin{figure}[here]
\centering
\includegraphics[width=12cm,height=10cm,keepaspectratio]{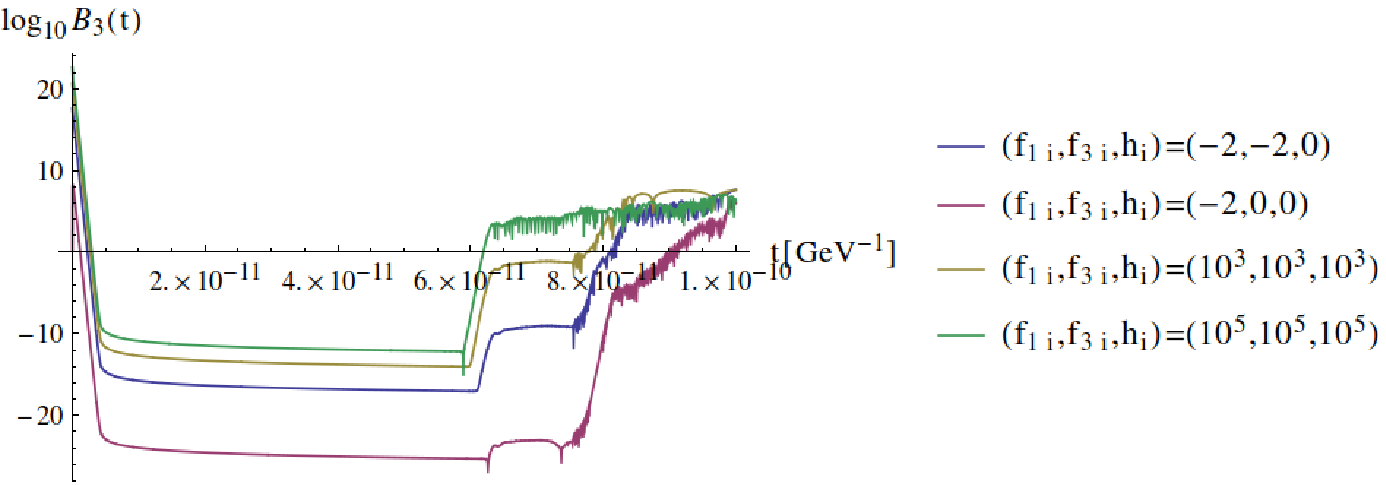}
\caption{The time evolution of $B_3(t)$ after inflation for several initial sets of gauge values with the initial conditions
 $\alpha_i=-20, \beta_i=0$.}
\label{fig: $logB3-2$}
\end{figure}%

\begin{figure}[here]
\centering
\includegraphics[width=12cm,height=10cm,keepaspectratio]{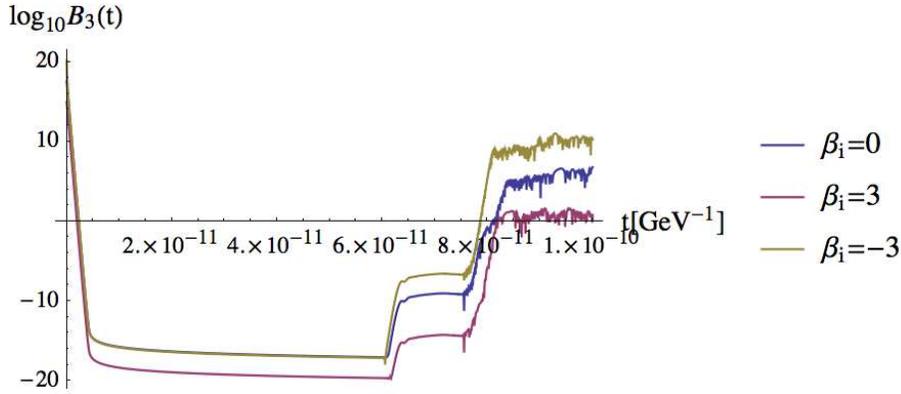}
\caption{The time evolution of $B_3(t)$ after inflation for several $\beta_i$ values with the initial conditions
 $f_{1i}=-2, f_{3i}=-2, h_{i}=0, \alpha_i=-20$. Apparently, the amplitude of magnetic fields depends on the anisotropy survived during inflation.}
\label{fig: $logB3-3$}
\end{figure}%

We depicted its evolution in time in Fig.\ref{fig: $logB3-2$}. Similar to the results in the previous section,
 the strength of magnetic fields also increase drastically after inflation. Remarkably, the final amplitude of magnetic fields
 depends on the initial anisotropy  as seen in Fig.\ref{fig: $logB3-3$}.  
This is because the anisotropy remaining during inflation affects the dynamics of the gauge fields.
We also notice there is a plateau where the growth due to parametric resonances balances with the decay due to the expansion of the universe.
At some point, however, the amplitude of magnetic fields starts to grow. It is intriguing to give an analytic understanding of this feature.

It would be interesting to see if the order of magnitude of magnetic fields generated by
 parametric resonances is comparable with observed magnetic fields. We expect the coherent oscillation of 
magnetic fields found in our study will be relevant when we consider inhomogeneous 
fluctuations of magnetic fields and lead to the observable magnetic fields with similar amplitudes.

 According to recent observations, the present magnetic field 
at cluster scales is expected in the range
\begin{equation}
10^{-15}[\text{G}] \lesssim B_{obs} \lesssim 10^{-9}[\text{G}] \ .
\end{equation}
Now, we would like to translate this range into  that after the reheating with a reheating temperature $T_{reh}$. 
Notice the relation $B_{reh}/B_{obs}=(a_{obs}^2/a_{reh})^2=(T_{reh}/T_{obs})^2$ and
 the present temperature $T_{obs}\sim10^{-4}[\text{eV}]=10^{-13}[\text{GeV}]$ 
determined by observations of CMB.  Then, using the relation $1[\text{G}]\simeq10^{-20}[\text{GeV}^2]$,  we obtain the range 
\begin{equation}
10^{11}\left(\dfrac{T_{reh}}{10^{10}[\text{GeV}]}\right)^2[\text{GeV}^2]\lesssim B_{reh} \lesssim 10^{17}\left(\dfrac{T_{reh}}{10^{10}[\text{GeV}]}\right)^2[\text{GeV}^2] \ .
\end{equation}
In the above, we used the following relationship (see for example \cite{Lyth:1998xn}) 
\begin{equation}
N_{COBE}=62-\text{ln}(10^{16} [\text{GeV}]/V_{end}^{1/4})-\dfrac{1}{3}\text{ln}(V_{end}^{1/4}/\rho_{reh}^{1/4}) \ ,
\end{equation}
where $V_{end}$ is the potential energy at the end of inflation, $\rho_{reh}$ is the energy density at the reheating time, 
and $N_{COBE}$  is  the e-folding number corresponding to the COBE observation.
Taking look at Fig.\ref{fig: $logB3-2$} and Fig.\ref{fig: $logB3-3$}, we see
magnetic fields in this range can be generated depending on initial conditions of $\beta$
or the reheating temperature $T_{reh}$.

\section{Conclusion}

 We studied the cosmological dynamics of the gauge fields in electroweak theory in the early universe. 
Since we have considered Higgs inflation, there exists non-trivial interaction between the inflaton and the gauge fields,
 which makes the difference from the previous work~\cite{Emoto:2002fb}. In spite of this difference, we found that there remains non-trivial configurations of the gauge fields during inflation  because of the presence of the flat direction.
 Main finding in this paper is the growth of the gauge fields due to parametric 
resonances induced by oscillations of the Higgs field.  
To characterize the topological configuration in the gauge fields, we calculated the Chern-Simons number and found the similar growth.
This implies that CP violating configuration of gauge fields exists, which would be relevant to  the baryogenesis. 
We have also estimated magnitude of magnetic fields
and found that the order of magnitude lies in the range of observed magnetic fields 
at cluster scales. Remarkably, the resultant amplitude of magnetic fields depends on
the anisotropy survived during inflation. This is an interesting manifestation of the anisotropy in Higgs inflation.  

In this work, we have not considered reheating process in detail.
To obtain precise results on magnetic fields, we need to introduce matter fields and
investigate reheating processes. Moreover, we have to study evolution of inhomogeneity 
during and after Higgs inflation. We leave these issues for future work.

\acknowledgements
I would like to thank Y. Fujimoto, Y.Sakakihara  and M.Watanabe for useful discussions. 
 This work was supported in part by the Grants-in-Aid for Scientific Research (C) No.25400251 
and Grants-in-Aid for Scientific Research on Innovative Areas No.26104708.

\end{document}